\documentclass[aip,reprint,superscriptaddress,showkeys]{revtex4-1}

\usepackage{graphicx,amsfonts,amsmath,amssymb}
\usepackage{soul}
\usepackage[latin1]{inputenc}
\usepackage[draft]{changes}
\usepackage{marginnote}

\newcommand{\mb}[1]{\mbox{\bfseries \itshape #1}}

\begin{document}

\title{Node differentiation dynamics along the route to synchronization 
in complex networks}

\author{Christophe Letellier}
\homepage{http://www.atomosyd.net/spip.php?article1}

\affiliation{
Rouen Normandie University --- CORIA, Campus Universitaire du Madrillet,
F-76800 Saint-Etienne du Rouvray, France }
\email{christophe.letellier@coria.fr}

\author{ Irene Sendi\~na-Nadal}

\affiliation{Complex Systems Group \& GISC, Universidad Rey Juan Carlos, 28933 
M\'ostoles, Madrid, Spain }
\email{irene.sendina@urjc.es}

\affiliation{Center for Biomedical Technology, Universidad Polit\'ecnica de 
Madrid, 28223 Pozuelo de Alarc\'on, Madrid, Spain}

\author{Ludovico Minati}
\affiliation{Center for Mind/Brain Sciences (CIMeC), University of Trento, 
38123 Trento, Italy}
\affiliation{Institute of Innovative Research, Tokyo Institute of Technology,
Yokohama, 226-8503, Japan}
\email{lminati@ieee.org}

\author{I. Leyva}
\affiliation{Complex Systems Group \& GISC, Universidad Rey Juan Carlos, 28933 
M\'ostoles, Madrid, Spain }
\affiliation{Center for Biomedical Technology, Universidad Polit\'ecnica de 
Madrid, 28223 Pozuelo de Alarc\'on, Madrid, Spain}
\email{inmaculada.leyva@urjc.es}

\date{\today}

\begin{abstract}
Synchronization has been the subject of intense research during decades mainly focused on determining the structural and 
dynamical conditions driving a set of interacting units to a coherent state globally stable. However, little attention has 
been paid to the description of the dynamical development of each individual networked unit 
in the process towards the synchronization of the whole ensemble. In this paper, we show how in a network of identical 
dynamical systems, nodes belonging to the same degree class differentiate in the same manner visiting a sequence of states 
of diverse complexity along the route to synchronization independently on the global network structure. In particular, we 
observe, just after interaction starts pulling orbits from the initially uncoupled attractor, a general reduction of the 
complexity of the dynamics of all units being more pronounced in those  with higher connectivity. In the weak coupling regime, when synchronization starts to build up, there is an increase in the dynamical complexity whose maximum is achieved, in general, first in the hubs due to their earlier synchronization with the mean field. For very strong coupling, just before complete synchronization, we found a hierarchical dynamical differentiation with lower degree nodes being the ones exhibiting the largest complexity departure. We unveil how this differentiation route holds for several models of nonlinear dynamics including toroidal chaos and how it depends on the coupling function. This study provides new insights to understand better strategies for network identification and control or to devise effective methods for network inference.

\end{abstract}
\keywords{Complex networks --- Dynamical complexity --- Route to 
synchronization --- Chaos}

\maketitle

\section{Introduction}

The description of networks can be structural, based on the characterization 
and modelling of their topological properties,\cite{New03,Boc06,Boc14} or 
dynamical, usually referring to the collective behavior emerging from the 
interaction between node dynamics and the network architecture. Synchronization 
is the most thoroughly investigated ensemble dynamics,\cite{Are06,Boc06,Rod16} 
including a rich variety of related behaviours such as chimera 
states,\cite{Kur02,Abr04,Hag12} cluster synchronization or \cite{Pec14} 
explosive synchronization.\cite{Gom11,Ley13} 

The route to the synchronous state has been approached from different points of 
view. Microscopically, it has been analyzed the different ways local 
synchronization grows as the coupling increases  depending on the connectivity 
structure.\cite{Are06,Res06,Gom07,Per20} Globally, the synchronizability of a 
population of identical networked oscillators, that is, the stability of the 
coherent state, has been tackled through the master stability function and 
foresees the emergence of patterns when the stability of the network uniform 
state is lost.\cite{Pec98,Bar02,Pec08} 

While these approaches help to understand how synchronization clusters grow and 
eventually merge into a macroscopic coherent state or predict its stability for 
a particular network structure and coupling function, few works pay attention 
to the description of the nodal dynamics along the route to synchronization.  
Recently, it has been shown that the degree of complexity of the node dynamics 
is strongly dependent on the connectivity; in particular, the local 
connectedness appears to confer considerable node differentiation hallmarks, 
albeit following mechanisms that are intricate and delineate a non-monotonic 
effect.\cite{Tla19a,Tla19,Min19} These ensemble fingerprints in 
the node dynamics can be used to infer the network statistical description, or 
even the detailed connectivity in certain conditions.\cite{Ero20} By node 
differentiation it is here meant as the dynamical departure of an isolated node 
due to the interaction with its neighborhood, being more pronounced in networks 
with nodes having degree heterogeneity. In particular, a relatively large range 
of coupling strengths over which nodes with a higher degree have a less 
complex dynamics has been reported,\cite{Tla19} explaining the low complexity 
observed in hubs of functional brain networks.\cite{Mar18} However, the 
reverse situation, the existence of an even lower range of coupling strengths 
where high degree nodes are more complex than lower degree ones, remains under 
study.\cite{Min19} The purpose of the present work is to elucidate the route of 
the node differentiation dynamics in a complex network before reaching a 
synchronous state, and, in particular, how it depends on the dynamical system, 
the coupling function and on the topological properties of the network 
structure. 

In order to characterize and quantify this route to node differentiation dynamics as the coupling strength increases, we 
used the complexity measure recently introduced by Letellier {\it et al}.\cite{Let20a} able to distinguish organized from 
disorganized chaotic behavior together with maps of the  dynamical patterns associated to relevant degree classes. Section \ref{sec:measure} is devoted to introducing terminology and providing a brief description of the dynamical complexity measure. We then investigated in Section \ref{sec:StarRoss} the dynamical differences among R\"ossler oscillators in a star network according to their degree, and how those differences dependent upon the coupling function, the nominal dynamics, and the size of the star. In Section \ref{sec:StarOther}, we extended the analysis to star networks of dynamical systems featuring higher nominal complexity, namely the symmetric Lorenz system, the high-dimensional Mackey-Glass equation, and the toroidal Saito system, to determine whether a general scenario can be outlined. In Section \ref{sec:largernets}, we considered R\"ossler systems in larger networks to explore the node differentiation induced by embedding the oscillators in a much more complex topological environment. Finally, Section \ref{conc} offers general conclusions.

\section{Dynamical complexity measure}
\label{sec:measure}

Let us start by considering the general description of a network comprising $N$ diffusively-coupled $m$-dimensional identical dynamical systems, whose state vector $\mb{x}_i$ evolves as
\begin{equation}
   \dot{\mb{x}}_i 
	= \mb{f}(\mb{x}_i) - d \sum_{j=1}^N\mathcal{L}_{ij}\,\mb{h}(\mb{x}_j)
   \label{eq:dynet}
\end{equation}
where $\mb{f}:  \mathbb{R}^m\rightarrow\mathbb{R}^m$ and $\mb{h}:\mathbb{R}^m\rightarrow\mathbb{R}^m$ are the nominal dynamics --- here, nominal refers to the vector field of the $i$-th nodal dynamics  when isolated --- and the coupling function, respectively, and $d$ is the coupling strength. $\mathcal{L}_{ij}$ are the elements of the Laplacian matrix encoding the network's connectivity, with $\mathcal{L}_{ii}=k_i$ the node degree, $\mathcal{L}_{ij}=-1$ if nodes $i$ and $j$ are connected, and $\mathcal{L}_{ij}=0$ otherwise. 
  
To characterize the nodal dynamics in a more reliable way, we computed  a Poincar\'e section to
the trajectory $\mb{x}_i(t)$ in the state space to rule out the local linear component and focus 
on  the  nonlinear signatures governing the dynamics.\cite{Let06b}  Thus, for each oscillator, in the following, we 
computed a two-dimensional Poincar\'e section of the attractor when the dynamics has a toroidal structure, or a 
first-return map built with one of the coordinates of the Poincar\'e section in case it is non-toroidal. Then, each 
oscillator's map is used to obtain a complexity coefficient $C_{\rm{D}}$ introduced in Ref.~\cite{Let20a} and defined as 
\begin{equation}
C_{{\rm D}} = S_{{\rm p}} + \Delta,    
\label{Cdef}
\end{equation}
being $S_{\rm p}$ a permutation entropy and $\Delta$ a structurality marker. We chose this complexity definition because of 
its ability to distinguish  organized unpredictable behavior (dissipative chaos) from disorganized unpredictable phenomena 
(noise or conservative chaos). While the entropy quantifies the unpredictability of the dynamics, the structurality 
accounts for its undescribability, two very different aspects contributing to the complexity of a dynamics. 

The permutation entropy $S_{\rm p}$ is defined as in Ref.~\cite{Ban02}:
\begin{equation}
  S_{\rm p} 
  = - \frac{1}{\log(N_{\rm s}!) } \sum_{\pi}^{N_{\rm s}!} p_{\pi} \log{p_{\pi}}
  \in [0,1] \, . 
\end{equation}
It is based on the probability distribution $p_{\pi}$ of the $N_{\rm s}!$ possible ordinal patterns constructed from the order relations of $N_{\rm s}$ successive data-points in the Poincar\'e section. The structurality is computed by dividing the first-return map (or the two-dimensional projection of the Poincar\'e section) into a $N_q \times N_q$ boxes and  
\begin{equation}
  \Delta = \sum_{i,j=1}^{N_q} \frac{q_{ij}}{N_q^2} \in [0,1] \, , 
\end{equation}
where $q_{ij}=1$ if the box $(i,j)$ is visited at least once, and $q_{ij}=0$ otherwise. The structurality $\Delta$ is typically low ($\Delta < 0.2$) for organized dynamics, and large ($\Delta > 0.80$) for disorganized dynamics. We thus have $C_{\rm D} \approx 0$ for a limit cycle ($S_{\rm p} \approx 0$ and $\Delta \approx 0$), $0.5 < C_{\rm D} < 1.0$ for dissipative chaotic behavior (organized but unpredictable), and $C_{\rm D} > 1.5$ for weakly dissipative chaos or stochastic processes (both are unpredictable and disorganized). 

In all our computations we used $N_s=6$ as the length of the ordinal patterns. The number $N_{\rm d}$ of data points in the 
Poincar\'e section to compute $S_{\rm p}$ and $\Delta$ was $N_{\rm d}=10,000$, largely above the requirement  
$N_{\rm d} > 5 N_{\rm s}!$ recommended by Riedl {\it et al.}\cite{Rie13} The number of boxes was 
$ N_q = 5 \log (N_{\rm d})$  chosen such that $N_{\rm d} \gg N_q$, as prescribed in Ref.~\cite{Let20a}. The box width 
$\delta_{\rm p}$ is determined by the largest range visited  in our simulations for a given system as 
$\delta_{\rm p} = (x_{\rm max} - x_{\rm min})/N_q$, where $x_{\rm max}$ and $x_{\rm min}$ are the maximal and minimal 
values recorded along one axis of the first-return map. To avoid excessively promoting noise contamination or small 
fluctuations around period-1 limit cycle, we introduced a ``noise filter'' to interpret the ordered patterns of the 
$N_{\rm s}$ data points to allow permutation only if $| x_i - x_j | > \delta_{\rm p}$.

\begin{figure}
  \centering
  \includegraphics[width=0.48\textwidth]{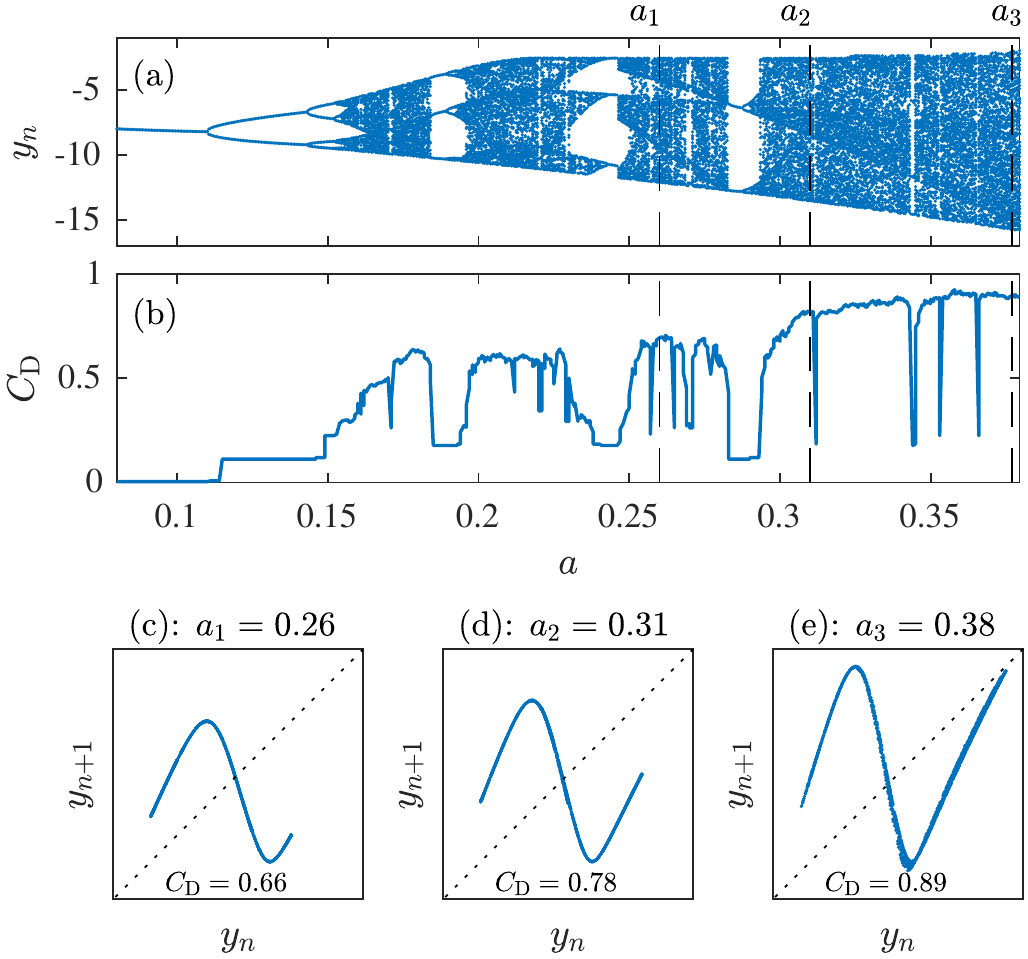} 
  \caption{Dynamical characterization of a single R\"ossler system upon 
variation of the $a$ parameter and fixing $b = 0.2$ and $c = 5.7$. Bifurcation 
diagram using the Poincar\'e section ${\cal P}_x$ (a) and dynamical complexity 
$C_{\rm D}$ (b). The dashed vertical lines in (a) and (b) are located at 
$a_1 = 0.26$, $a_2=0.31$, and $a_3=0.38$. The corresponding first-return maps from 
$y_n$ are shown in (c)-(e) and their complexity values at the bottom of each panel.}
  \label{fig1}
\end{figure}

Let us consider the paradigmatic R\"ossler system\cite{Ros76c} to exemplify the ability of Eq.(\ref{Cdef}) to discriminate between different dynamics. The vector flow in  Eq.~(\ref{eq:dynet}) is
\begin{equation}
  \mb{f}(\mb{x}) = \left[-y-z,x+a y,b+z(x-c)\right]  
\end{equation}
taking $b=0.2$ and $c=5.7$ as fixed parameters and $a \in [0.08,0.38]$ as the 
bifurcation parameter. Equation (\ref{eq:dynet}) was integrated using a 
fourth-order Runge-Kutta scheme \cite{Pre92} with a time step 
$\delta t = 0.01$. Initial conditions are randomly selected within a small 
neighborhood of radius $0.5$ centered at the origin of the state space. 
Figure~\ref{fig1}(a) shows the bifurcation diagram as a function of the parameter $a$ for a single R\"ossler system ($d=0$)
computed from the $y_n$ coordinate of the Poincar\'e section  
\begin{equation}
  {\cal P}_x \equiv \left\{ \displaystyle (y_n, z_n) \in \mathbb{R}^2 ~|~   x_n = x_-, \dot{x}_n > 0 \right\} \, , 
\end{equation}
where $x_- = (c - \sqrt{ c^2 - 4 ab})/2$ is the coordinate of the inner 
singular point.\cite{Let95a} The classical period-doubling cascade as a route to chaos is 
quantitatively characterized by the dynamical complexity $C_{\rm D}$ in Fig.~\ref{fig1}(b). It  perfectly captures the periodic windows 
($C_{\rm D} < 0.2$) intermingled with chaotic behavior. Panels (c)-(e) show the first-return maps for the three $a$ values marked in (a)-(b) with vertical lines. The first value, $a=0.26$ is located between a period-3 and a period-2 window and its first-return map is bimodal [Fig.~\ref{fig1}(c)] but with a slightly developed third branch appearing just after the period-3 window. For $a=0.31$, Fig.~\ref{fig1}(d), the bimodal map has the three branches. A more developed chaos, 
characterized with three monotone branches, is observed for $a = 0.38$ in
Fig.~\ref{fig1}(e). In particular, the right end of the third 
branch approaches the vicinity of the bisecting line, indicating that a new period-1 orbit is about to be created in the population of unstable periodic orbits. This is the most 
developed chaos that can be observed along this line of the parameter space, as indicated by the dynamical complexity ($C_{\rm D} = 0.89$), in agreement with 
the most developed first-return map.

\begin{figure}[t]
  \centering
  \includegraphics[width=0.49\textwidth]{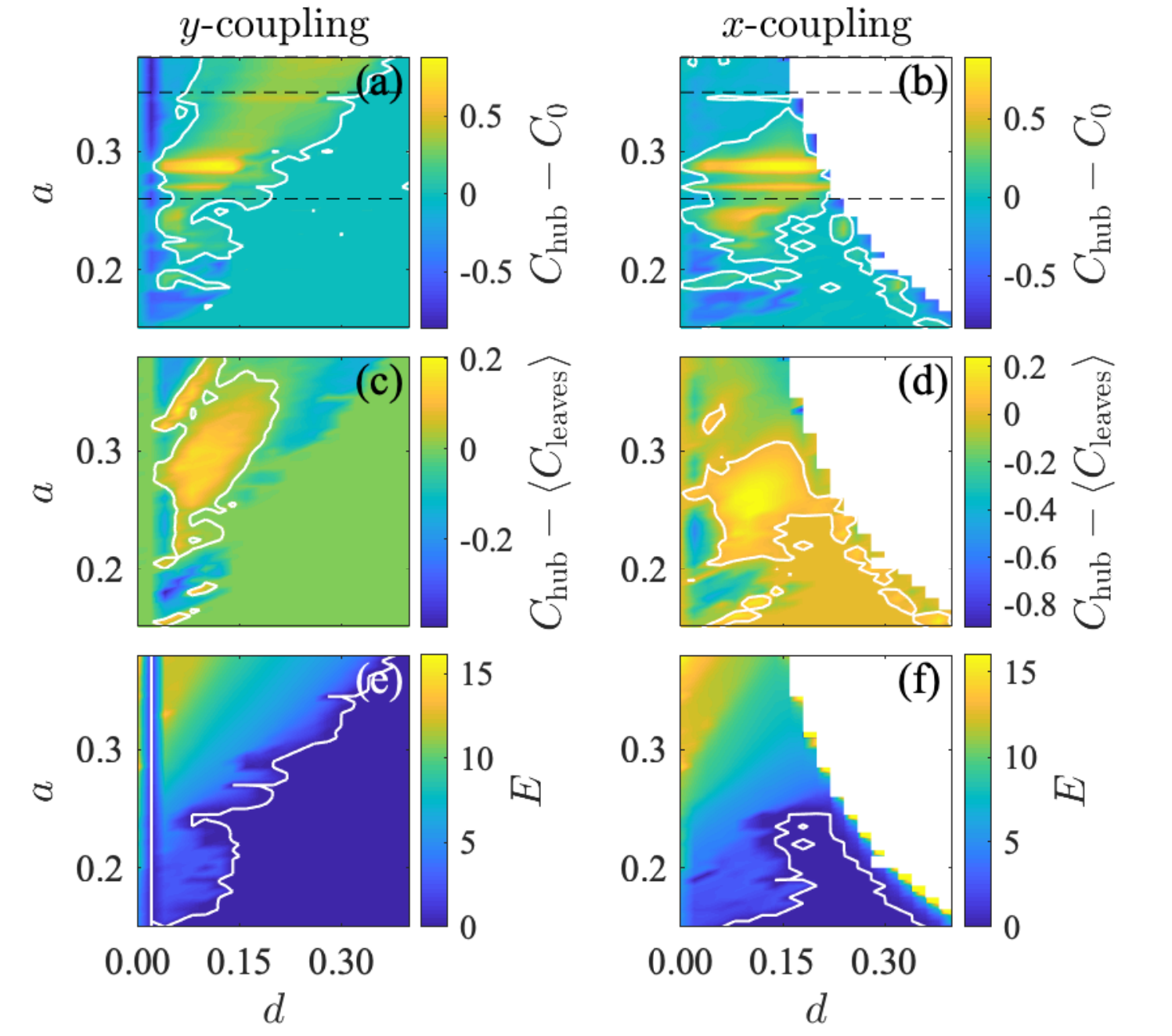}  \\[-0.3cm]
  \caption{Colormaps of the complexity values for a star of $N=16$ coupled 
R\"ossler oscillators in the $d$-$a$ parameter plane.
Actual complexity change of the hub $C_{\rm hub} - C_0$  (a)-(b) and the 
difference $C_{\rm hub} - \langle C_{\rm leaves} \rangle$ 
(c)-(d) when oscillators are coupled through the $y$ (a,c), and $x$ (b,d)
variables, respectively. Bottom panels (e)-(f) show the synchronization error $E$. The color scales are provided on the right sides.  In (a)-(f) panels, white curves are the corresponding null isolines, and horizontal dashed lines in (a)-(b)  are plotted at $a=0.26, 0.31$ and $a=0.38$.
White domains  correspond to the ejection of the trajectory to infinity. Other parameter 
values are set as in Fig.~\ref{fig1}.}
  \label{fig2}
\end{figure}

\section{A star network of R\"ossler systems}
\label{sec:StarRoss}

Let us now consider a star network of $N$ R\"ossler systems coupled according to  Eq.~(\ref{eq:dynet}), with $N-1$ peripheral nodes and one central 
node acting as the hub, aiming at exploring the influence of the coupling function in the parameter space $d-a$. Top panels (a) and (b) of Fig.~\ref{fig2} show the  actual change in complexity $C_{\rm hub} - C_0$ of the hub in a $N=16$ star with respect to the reference complexity value $C_0$ (which is a function of the parameter $a$) of an uncoupled node, for coupling schemes through variables $y$  and
$x$, respectively. Middle panels (c) and (d) provide the complexity difference between the hub and the leaves $C_{\rm hub} - \langle C_{\rm leaves} \rangle $.

From the colormaps, the first clear observation is that the coupling function 
has a role in the node differentiation dynamics, as reflected by the different distribution of complexity change on the $a$-$d$ plane, which strongly depends on the network synchronizability. Panels (e,f) in Fig.\ref{fig2} show the time-averaged synchronization error computed as 
\begin{equation}
    E = \frac{2}{N (N-1)} 
    \displaystyle \sum_{i \ne j} \| \mb{x}_i-\mb{x}_j \| \, .
\end{equation}
which is in agreement with the prediction given by the master stability function (MSF) for R\"ossler systems coupled through the $y$ variable, type I synchronizability class -the MSF becomes negative above a critical value, and $x$ variable, type II class -the MSF is negative in a bounded interval.\cite{Pec98,Hua09,Sen16}
For the particular network structure of a star of size $N=16$, complete synchronization is indeed reached for a larger number of pairs 
($a$,$d$) when R\"ossler systems are $y$-coupled than when $x$-coupled. 
In the latter case, full synchronization cannot be reached for $a\gtrsim 0.25$ before a boundary crisis ejects the trajectory to infinity (white domains in Fig.\ \ref{fig2}(f)). For this configuration, the location of $d\lambda_i$, being $\lambda_i$ the $i$-th eigenvalue of the Laplacian matrix, is outside the range where the MSF is negative for any value of $d$, and the synchronous solution is always unstable. 

Roughly, in the regions where the synchronization error is not null, for both $x$ and $y$ coupling, there is a large region 
of the parameter space with yelowish areas delimited by the white null isolines in panels in Figs. \ref{fig2}(a) and 
\ref{fig2}(b) where the effect of coupling is to render the dynamics of the hub more complex with respect to its uncoupled 
regime ($C_{\rm hub}>C_0$). Outside this region, there are smaller islands (in blue) where the dynamics turns out to be 
less complex. However, when comparing this departure from the uncoupled dynamics between the hub and the leaves, Figs. \ref{fig2}(c) and \ref{fig2}(d), there are regions where the hub dynamics appears to be more complex than that of the 
leaves  --- for intermediate 
values  of the coupling and of the parameter $a$ --- and regions where the opposite is realized. For example, for the $y$ coupling, $C_{\rm hub}<\langle C_{\rm leaves} \rangle$ at the left and right side of the yellow region.

\begin{figure}
  \centering
  \includegraphics[width=0.37\textwidth]{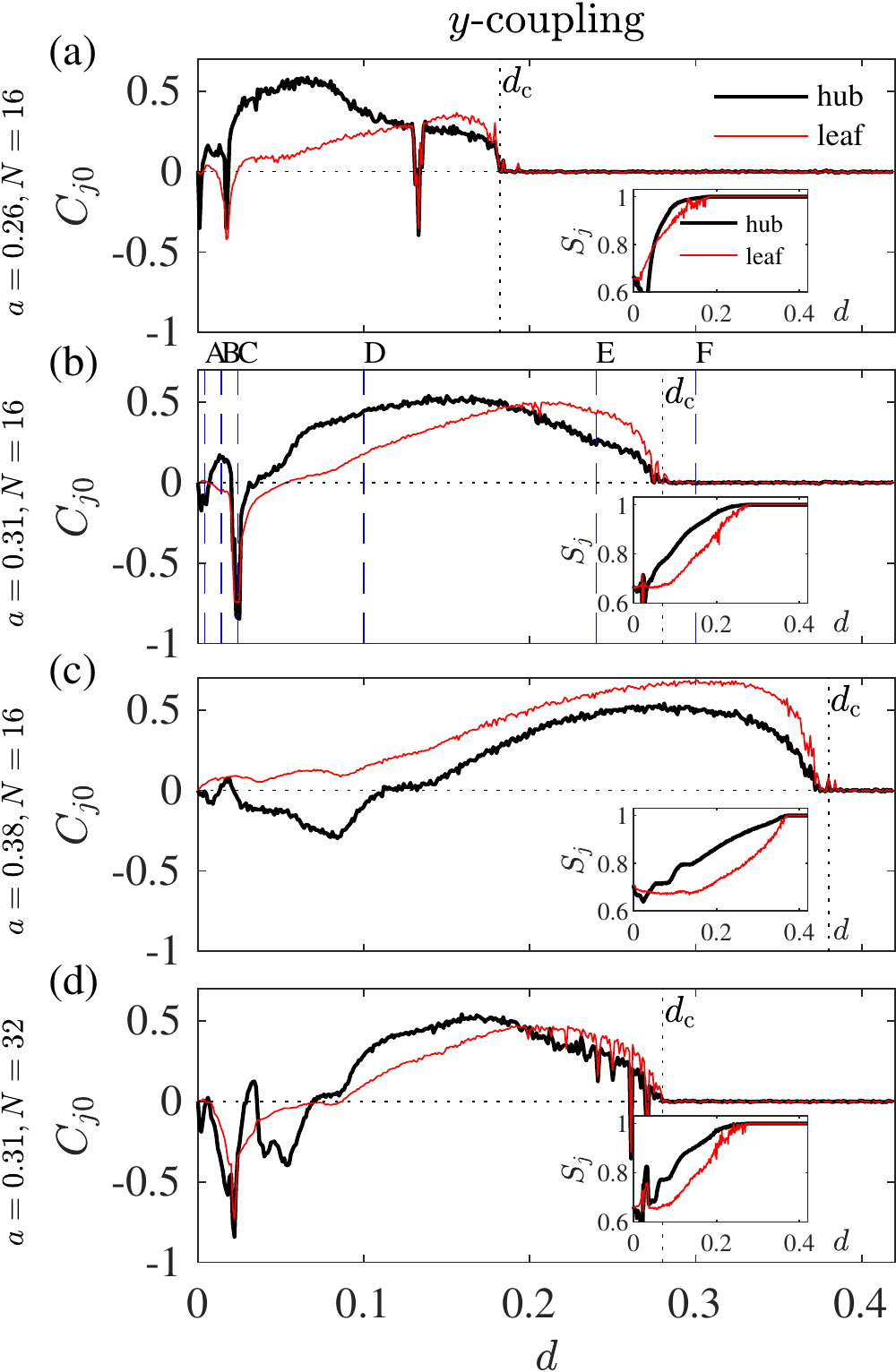} \\[0.1cm]
   \includegraphics[width=0.37\textwidth]{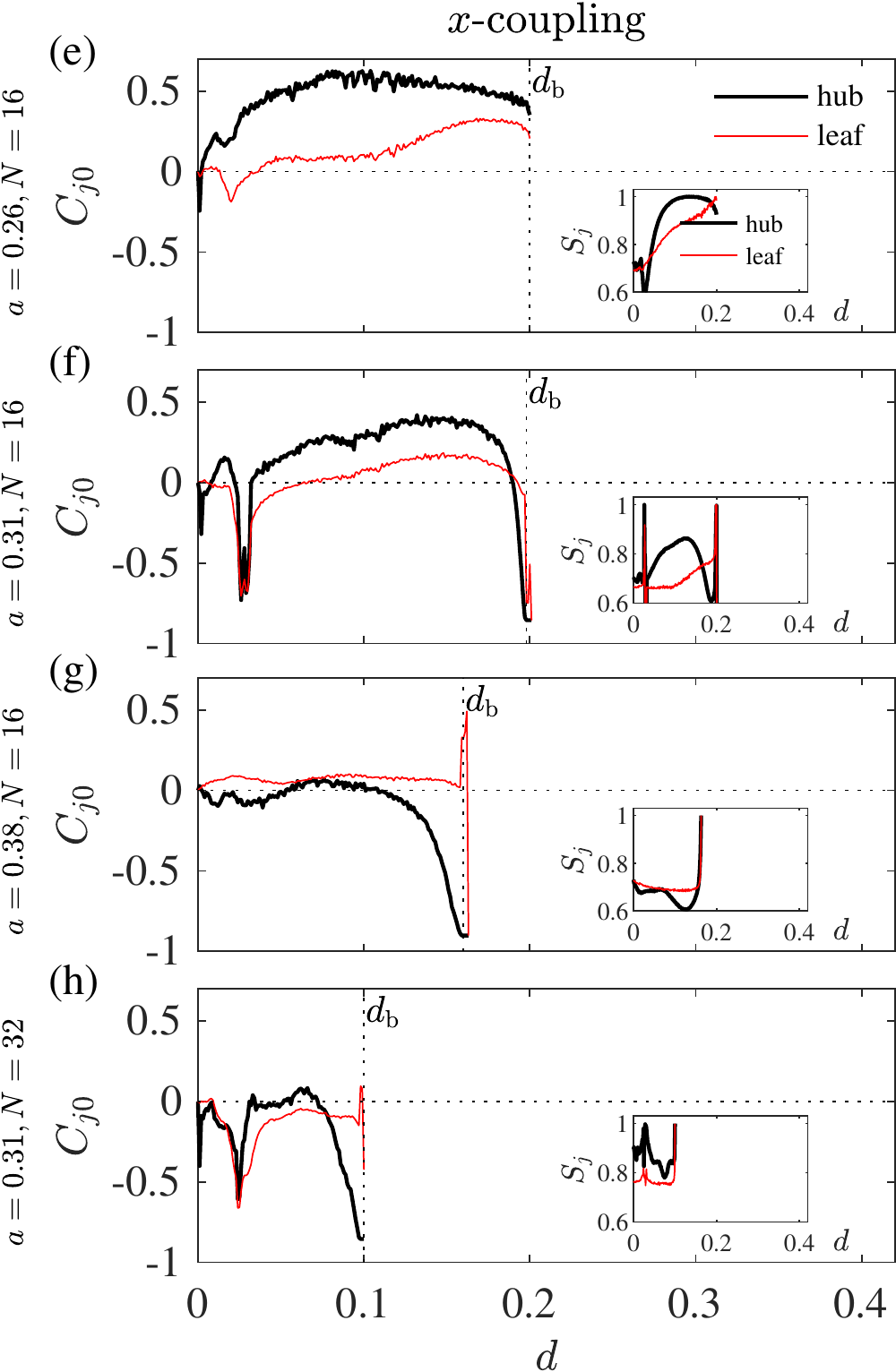} \\[-0.2cm]
  \caption{Evolution of the relative complexity $C_{j0}$  as a function of the 
coupling strength $d$ for the hub (in black) and 
a leaf (in red) of a star network of  $y$ coupled  (a)-(d) and $x$ coupled (e)-(h) R\"ossler oscillators. (a,e) $a=0.26$, (b,f) $a=0.31$, and (c,g) $a=0.38$ and $N=16$. (d,h) $a=0.31$ and $N=32$. Insets in all panels show the phase synchronization $S_j$ of the hub (black) and the leaf (red) with the mean field.  Other parameter values: $b=0.2$, and $c = 5.7$. One run	per $d$-value. }
  \label{fig3}
\end{figure}

To analyse with further detail the dynamical differentiation of hub and leaves along the route to synchronization, we show 
in Fig.~\ref{fig3}  cuts of the complexity difference $C_{j0}=C_{j}-C_0$, with $ j = \{ {\rm hub, leaf}\}$ 
(black for the hub and red for one leaf) along the three horizontal dashed lines shown in Fig. \ref{fig2}(a).
These curves clearly show that hub and leaf experience unalike differentiation routes before either the star synchronizes at $d_{\rm c}$ to the same dynamical state as in $d=0$ for $y$-coupling [Fig.~\ref{fig3}(a)-(d)] or it reaches a boundary crisis at $d_{\rm b}$ for the $x$ coupling [Fig.~\ref{fig3}(e)-(h)]. In all cases, the maximal differentiation occurs always before for the hub than for the leaf, that is, the maximum of the $C_{j0}$ curves is located at a coupling that is lower for the hub than for the leaf. As it is shown at the inset of each panel, this shift coincides with the earlier synchronization of the hub to the mean field, here measured as 
 \begin{equation}
  	S_{j} = \displaystyle \langle
  	\operatorname{Re}(e^{{\rm i}(\theta_j-\Phi)})\rangle_t\, , 
\label{S_j}
  \end{equation}
  being $\langle\dots \rangle_t $ the time average, $\operatorname{Re}$ stands for the real part, $\theta_{j} = \arctan \left(y_{j}/x_{j} \right)$ the phase of the $j={\rm hub/leaf}$, and $\Phi$ the mean-field global phase. By definition, $S_{j}\in[0,1]$, and approaches $1$ as the phase of the $j$-th oscillator is locked to the phase of the mean field.

\begin{figure*}
	\includegraphics[width=0.90\textwidth]{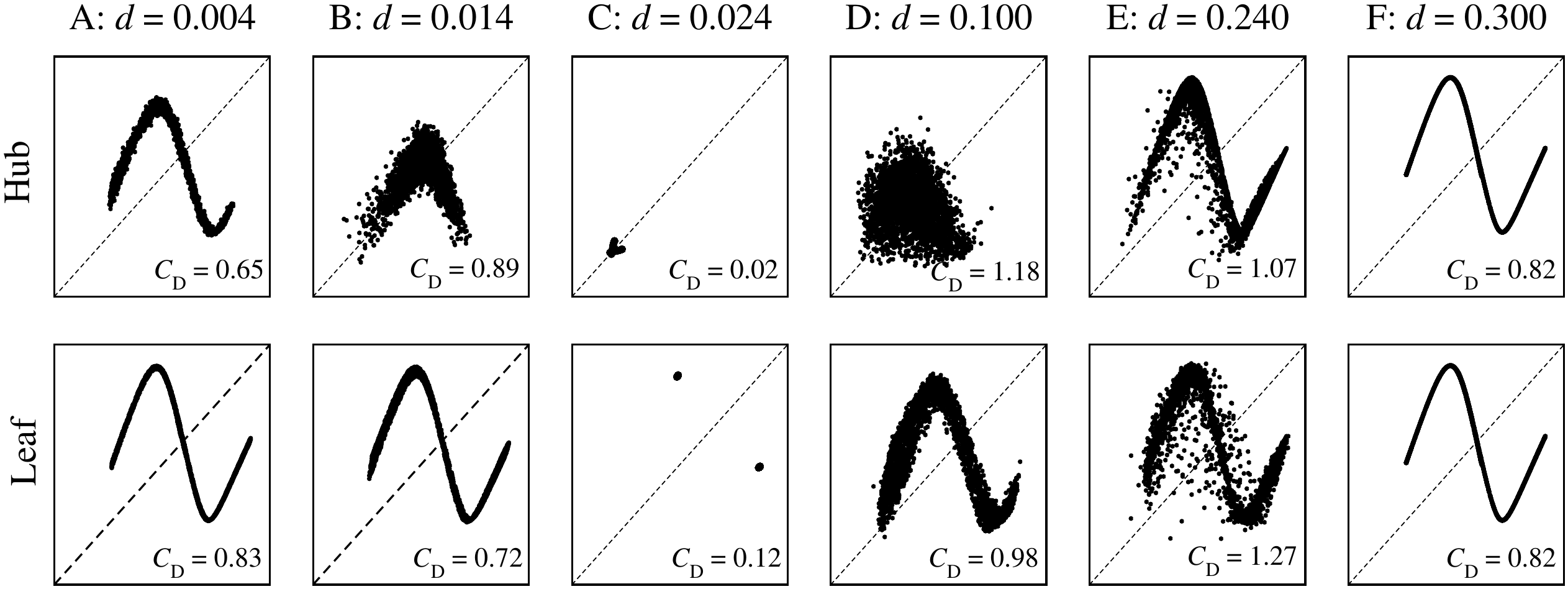}
  \caption{First-return maps to the Poincar\'e section ${\cal P}$ for the hub 
(top row) and one of the leaves (bottom row) in a star network of 
$N = 16$ $y$-coupled R\"ossler oscillators for $a = 0.31$ and for different $d$ 
values. The corresponding complexities $C_{\rm D}$ are reported. The nominal 
complexity is $C_0 = 0.82$.}
  \label{fig4}
\end{figure*}

Another observed systematic behavior is that, for increasing values of  $a$, as the uncoupled dynamics is more developed (see Fig.~\ref{fig1}), the relative complexity of the hub diminishes while increases for the leaf. Therefore, the relative position between the hub and leaf curves changes such that for $a=0.26$ the hub's curve is almost always above the leaf's one, $C_{\rm hub,0}>C_{\rm leaf,0}$, while  for $a=0.38$, when the isolated dynamics is the most developed, the hub's curve is always below,  $C_{\rm hub,0}<C_{\rm leaf,0}$. Another salient feature is that in the very weak coupling regime ($d<0.05$), just after oscillators start interacting, there is a marked fall of the complexity with respect to $C_0$ for both hub and leaf. 
Finally, the impact of the size of the star is analysed in Fig.~\ref{fig3}(d,h) for $N=32$ and $a=0.31$ which has to be compared with the panels (b) and (f) of the same figure for $N=16$. Precisely, the hub complexity is slightly affected and only in the region of low coupling. As predicted by the MSF, for the $y$ coupling, the threshold for synchronization $d_c$ does only depend on the smallest non-zero eigenvalue $\lambda_2$, which  equals $\lambda_2=1$ in both configurations. On the other hand, for the $x$ coupling, the boundary crisis is anticipated to smaller coupling values as the largest eigenvalue $\lambda_N=N$ increases with the size.

To picture the different dynamical states the hub and the leaf are visiting, we monitored their first-return maps at a specific points  along the route to synchronization as shown in Fig.~\ref{fig3}(b) with vertical dashed lines. 
Having in mind the map characterizing the uncoupled dynamics for $a=0.31$ and shown in Fig.~\ref{fig1}(d), for very low coupling [Fig.\ \ref{fig4}(A)], the leaf dynamics is nearly unaffected while the hub is exhibiting the presence of incoherent small perturbations and a shorter third branch. 
A slight increase in the coupling strength [Fig.\ \ref{fig4}(B)] amplifies this effect turning the hub's attractor into a very thick unimodal map and leaving the leaf's map almost unaltered. 
The next scenario [Fig.\ \ref{fig4}(C)] corresponds to the largest deviation from the uncoupled dynamics with both types of nodes displaying a much less complex dynamics. The hub is constrained to a small neighborhood of the inner
singular point as revealed by the feeble cloud of points in the bisecting line. The leaf is locked on a period-2 limit cycle. Beyond this drop of complexity, 
the hub dynamics develops into a single huge cloud of points, displaced along the bisecting line towards the period-1 limit cycle [Fig.\ \ref{fig4}(D)]. The large complexity value attained by the hub $C_{\rm hub} = 1.18 >1$ indicates a very disorganized dynamics far from the typical noisy limit cycle. At the same time, the leaf moves into a first-return map with three thick branches due to the noisy feedback from the hub. Pursuing along the route to synchronization both hub and leaf each recover three branched maps slightly distorted until a synchronous state is recovered for $d>d_c$ in Fig.\ 
\ref{fig4}(F) and all nodes share the same dynamics as they had when uncoupled. 

\section{Star networks made of other dynamical systems}
\label{sec:StarOther}

\subsection{Lorenz systems}

\begin{figure}
  \centering
	\includegraphics[width=0.48\textwidth]{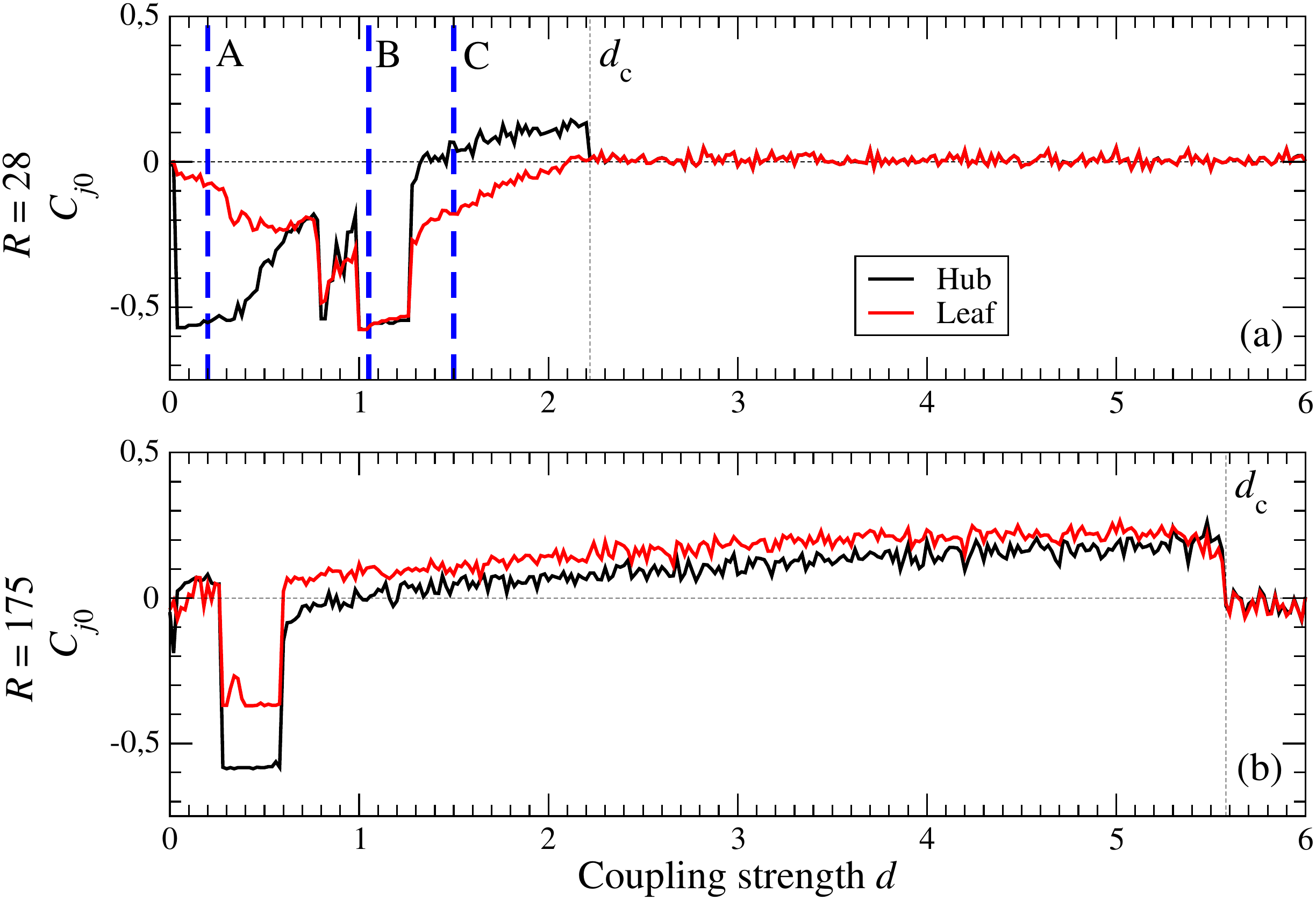} \\[0.2cm]
  \includegraphics[width=0.45\textwidth]{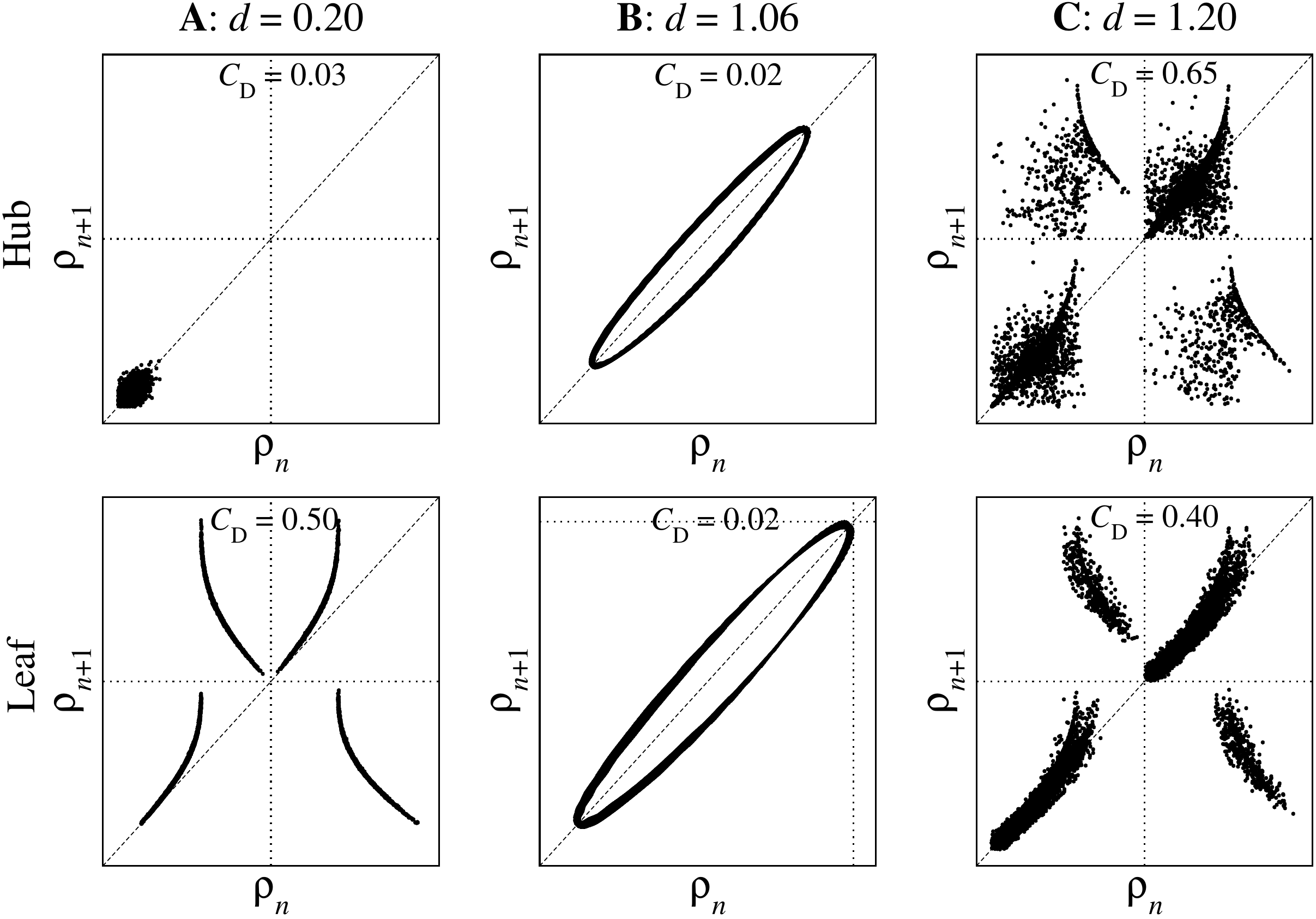} \\[-0.3cm]
  \caption{(a)-(b) Relative complexity for the hub (black) and one of the leaves 
(red) as a function of the coupling strength $d$ for a star network with $N=32$ 
$y$-coupled Lorenz systems for two $R$-values. First-return maps of the 
hub (top) and of one leaf (bottom) for the cases A, B and C indicated by the 
blue dashed lines in the top panel. Other parameter values: $\sigma = 10$, 
$b = \frac{8}{3}$.  A magnified view of the 
maps is plotted for the case B: the range used is not the same between the hub and leaf.}
 \label{fig5lor}
\end{figure}
  
Let us now consider the Lorenz 63 system \cite{Lor63} whose vector flow in Eq.~(\ref{eq:dynet}) is 
\begin{equation}
  \mb{f}(\mb{x}) = \left[\sigma (y -x),Rx -y -xz, -bz + xy   \right]  
\end{equation}
which is equivariant under a 
rotation symmetry around the $z$-axis\cite{Let01}. This global property  is the main difference
with respect to the R\"ossler dynamics since, when the symmetry is modded out, the Lorenz 
attractor is topologically equivalent to the R\"ossler one.\cite{Let94a,LetPhD,Let01} Lorenz systems coupled through variable $y$ present a type-{\sc I} synchronizability class and, therefore, complete synchronization is stable above a critical coupling threshold.

To compute the first-return map we proceeded as follows. The common Lorenz attractor for $R=28$, resembling the two wings of a butterfly, is bounded by a genus-3 torus and the Poincar\'e section is the union of the two components
\begin{equation}
  {\cal P}_\pm \equiv
  \left\{
    \left( \displaystyle y_n, z_n \right) \in \mathbb{R}^2 ~|~
	x_n = \pm x_\pm, \dot{x}_n \lessgtr 0 
	\right\}
\end{equation}
where $x_\pm = \pm \sqrt{b (R-1)}$.\cite{Let94a,Tsa04,Let05a} The interval  visited by each one of these 
components is normalized to the unit interval, and the component ${\cal P}_-$ is shifted by $-1$, 
leading to variable $\rho$. 
A quite close example of the first-return map for $R=28$ is shown at the bottom of Fig.~\ref{fig5lor}A 
featuring four branches paired  due to the rotation symmetry.\cite{Byr04}
The two increasing branches correspond to the reinjection of the trajectory into the wing from which it is issued, while the two decreasing branches are associated with the transition from one wing to the 
other. The two left (right) branches correspond to the $n$th intersection in the 
left (right) wing and to the ($n+1$)th intersection in the left (right) or 
right (left) wing depending on the sign of the slope.

Figure \ref{fig5lor} shows the relative complexity $C_{ j0}$ of the hub and a leaf of a $N=32$ star network $y$-coupled 
Lorenz systems for two values of the parameter $R$. For $R=28$, full synchronization is reached for 
$d_{\rm c} \approx 2.22$.  As with the R\"ossler systems, the complexity of the hub and leaves drop below $C_0$ in the 
weak coupling regime ($d\lesssim 1$) and beyond that point 
the hub becomes more complex than the nominal dynamics before full synchronization is eventually reached. 
For this parameter setting, the Lorenz hub experiences a sudden drop in complexity as illustrated in the first-return map [Fig.\ \ref{fig5lor}(A)]: 
while the hub dynamics is characterized by small fluctuations around the singular point in the centre of one of the wings, the leaves are nearly unperturbed with their four branches. When $d$ is further increased [Fig.\ \ref{fig5lor}(B)], the leaf dynamics also collapses, with all nodes surprisingly exhibiting a quasi-periodic dynamics, a regime not observed in an isolated Lorenz system. Notice that the size of the two tori are different. When the nodes' dynamics have topologically equivalent attractors (here, tori) but with a scaling 
factor, the phenomenon is known as {\it amplitude enveloppe 
synchronization}.\cite{Gon02} Finally, beyond this point of reduced dynamics and just before full synchronization [Fig.\ \ref{fig5lor}(C)], the interaction between hub and leaves moves the dynamics of the leaves again to the four paired branches although slightly more noisy than the uncoupled one, while the hub is strongly perturbed with a very disorganized dynamics. This scenario resembles the one observed for the R\"ossler system depicted in panel D of Fig. \ref{fig4}.
Finally, we explored a second regime with an even more developed uncoupled dynamics for $R = 175$ whose first-return map has six branches (not shown). 
As in Fig. \ref{fig3}(c) for the R\"ossler system, the switch between the relative complexities of hub and leaf is no  
longer observed in Fig. \ref{fig5lor} and the leaf curve is always above 
$C_{\rm leaf,0} > C_{\rm hub,0}$.

\subsection{Mackey-Glass delay differential equation}

\begin{figure}
  \centering
  \includegraphics[width=0.48\textwidth]{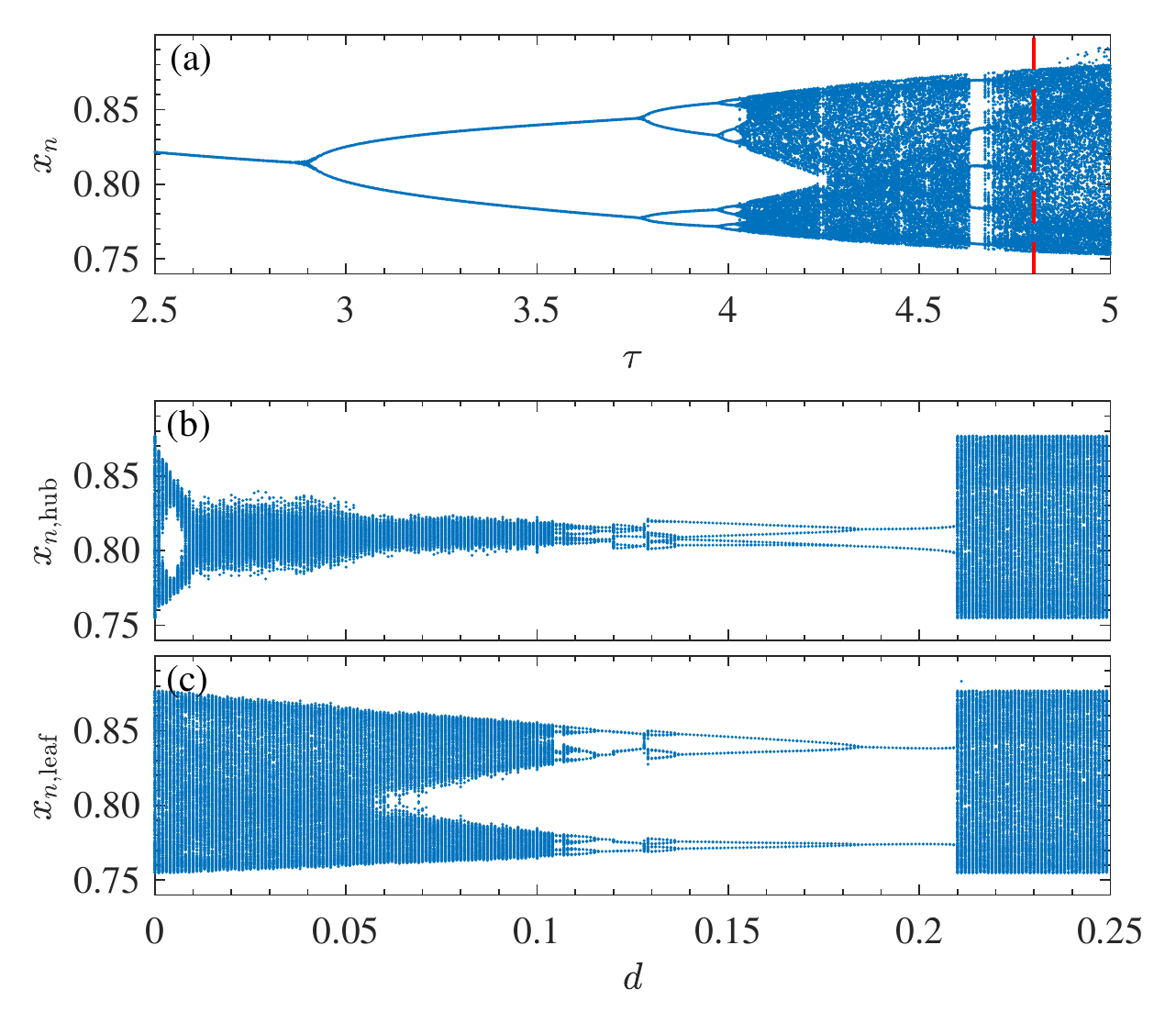} \\[-0.5cm]
  \caption{(a) Bifurcation diagram for the Mackey-Glass delay differential 
equation (\ref{macgla}) versus the time delay $\tau$. (b-c) Bifurcation 
diagrams computed versus the coupling strength $d$ for the hub and one of the 
leaves of a star network of $N = 32$ coupled Mackey-Glass equations 
(\ref{macgla}) for $\tau=4.8$ [see red-dashed vertical line in panel (a)]. For 
avoiding the multi-stability which is observed in such a network, the 
bifurcation diagram is computed without a reset of the initial condition at 
each new value of the coupling strength. Other parameter values are  
$\mu = 1.2$ and $p= 18.5$.}
  \label{fig7MKGbifs48}
\end{figure} 
It is possible to produce a chaotic attractor characterized by a smooth unimodal
map via one-dimensional delay differential equations, for instance, the 
Mackey-Glass (MG) equation\cite{Mac77,Gla79}
\begin{equation}
   \label{macgla}
   \dot{x} = \mu \frac{x_\tau }{1 + x_\tau^p} - x_t
 \end{equation}
where $x$ is the population of blood cells, $x_t = x(t)$ and 
$x_\tau = x(t - \tau)$. This equation was initially proposed for the control of 
hematopoiesis (the production of blood cells). Typically, the delay $\tau$ is the time-scale for proliferation and maturation of these blood cells. The dimension of the effective state space associated with a delay differential equation is dependent on the delay.\cite{Gum74,Far82} Parameter values for parameters $\mu$ and $p$ are such that a period-doubling cascade as a route to chaos is observed [Fig.\ \ref{fig7MKGbifs48}(a)]. We 
chose a delay $\tau=4.8$ (red dashed line in Fig.\ \ref{fig7MKGbifs48}(a) whose attractor is characterized by 
a smooth unimodal map with equivalent topological properties to the R\"ossler system. Increasing $\tau$ further leads to a much more complex behavior and computing a reliable Poincar\'e section is rather tricky. 

Our goal here is to investigate the route to synchronization for unimodal dynamics produced by a potentially high-dimensional system. A network of MG systems can be fully synchronized using bidirectional coupling.\cite{Pyr98} 

\begin{figure}
  \centering
  \includegraphics[width=0.48\textwidth]{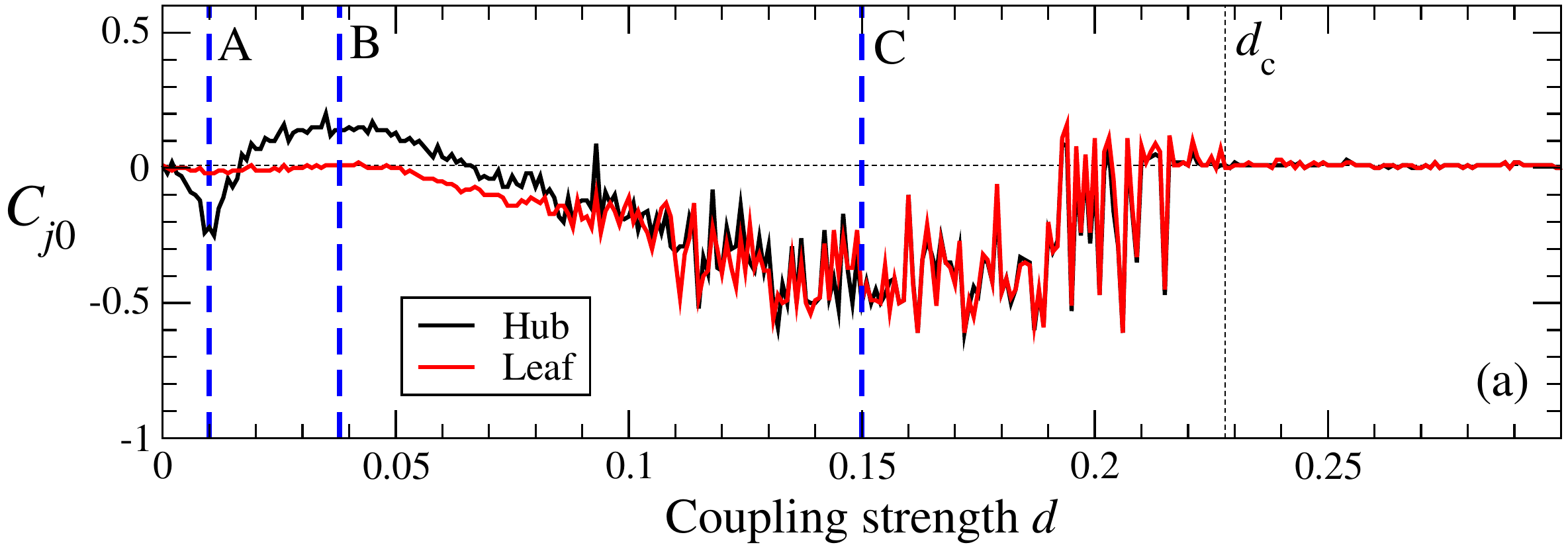} \\[0.2cm]
  \includegraphics[width=0.45\textwidth]{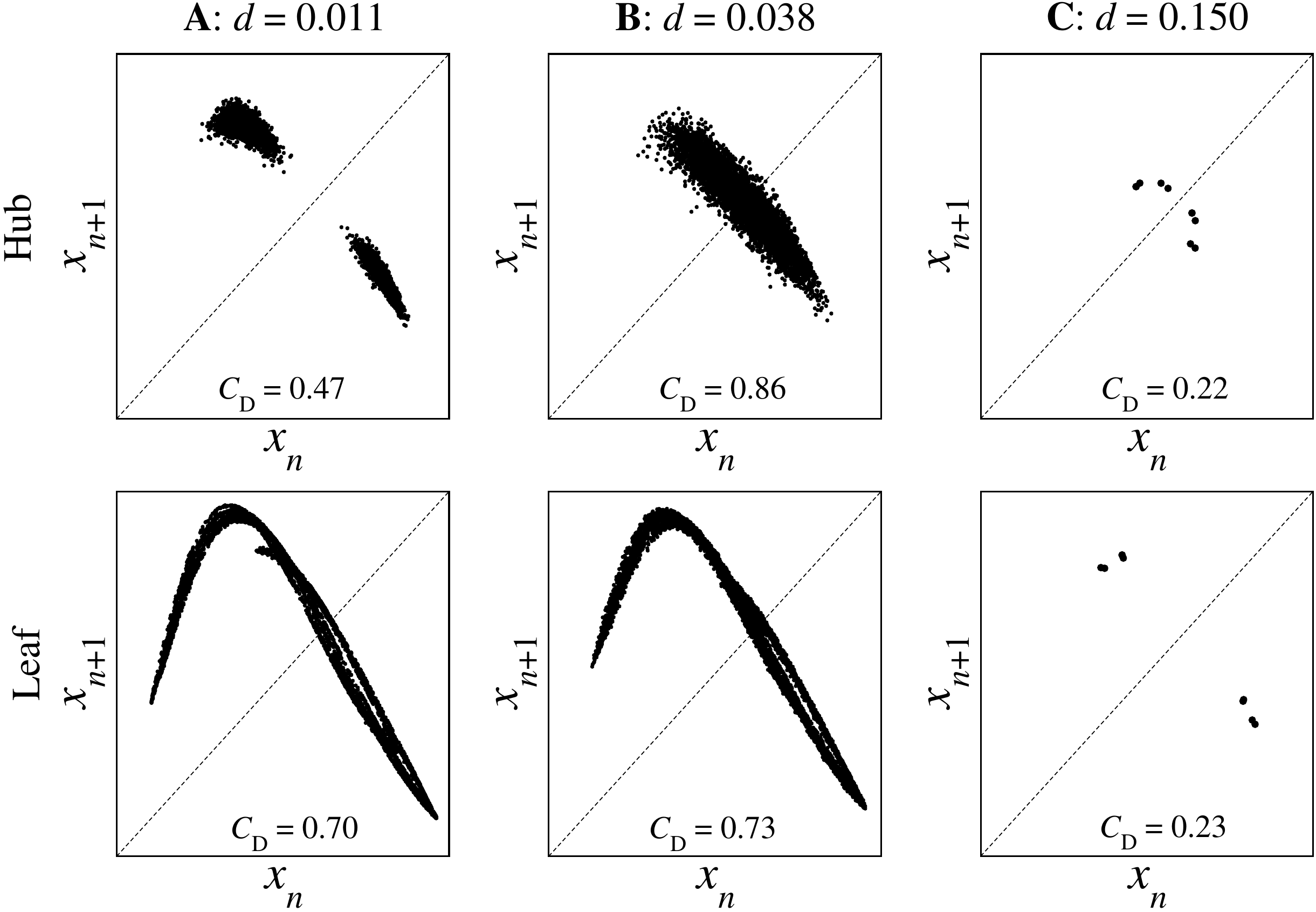} \\[-0.3cm]
  \caption{(a) Relative complexity as a function of the coupling strength $d$ for a  star network of 
$N=16$ coupled Mackey-Glass delay differential equations (\ref{macgla}) for $\tau=4.8$. (A)-(C) First-return 
maps of the hub (top) and of  one leaf (bottom) for three different values of $d$. Other parameter values: 
$\mu = 1.2$ and $p= 18.5$.}
 \label{MKGmaps3}
\end{figure}

The attractor is bounded by a genus-1 torus and the single-component Poincar\'e section can be defined as 
\begin{equation}
  {\cal P}_{\rm MG} \equiv
  \left\{
     x_n, \in \mathbb{R} ~|~ \dot{x}_n  = 0.025, \ddot{x}_n > 0, x_n < 0.9
  \right\} \, . 
\end{equation}
The first-return map is one-dimensional and slightly foliated and quite similar to the one shown for the leaf
in Fig.\ref{MKGmaps3}(A) for $N=16$ coupled MG systems. Figure \ref{MKGmaps3}(a) shows the relative complexity 
of the hub and leaf as a function of the coupling strength. As observed in the R\"ossler and Lorenz systems, 
for low $d$-values, the hub experiences a sudden complexity drop while the leaves keep almost their nominal 
dynamics. As the coupling is increased, the hub starts synchronizing with the mean field and its relative 
complexity rises above $0$. Beyond point B, all nodes reduce their complexity down to a  minimal value 
(point C) before finally increasing up to full synchronization is reached. The noisy curves in that region is 
due to the extreme sensitivity to initial conditions found in this system. The  maps for the  hub and leaf 
[Figs.\ \ref{MKGmaps3}(A)-\ref{MKGmaps3}(C)] corresponding to the setting points A,B, and C marked in Fig.\ \ref{MKGmaps3}(a) sketch the dynamical differentiation route. 

The bifurcation diagrams for the hub and leaf respectively are computed as a function of the coupling $d$
[Figs. \ref{fig7MKGbifs48}(b)-\ref{fig7MKGbifs48}(c)]. In both cases, an inverse period-doubling cascade is observed up to a period-2 limit cycle is settled at $d = 0.18$. Curiously, the size of both limit cycles is different which, again, is an example of an  amplitude envelope synchronization. The crisis leading to a larger chaotic attractor is strongly dependent on the initial conditions. 
In summary, the prevalent lines of the route to synchronization  previously sketched are visible observed, albeit with some differences. Further investigations are necessary to fully understand their origin.

\subsection{Four-dimensional Saito model}

 \begin{figure}[t]
   \centering
   \includegraphics[width=0.48\textwidth]{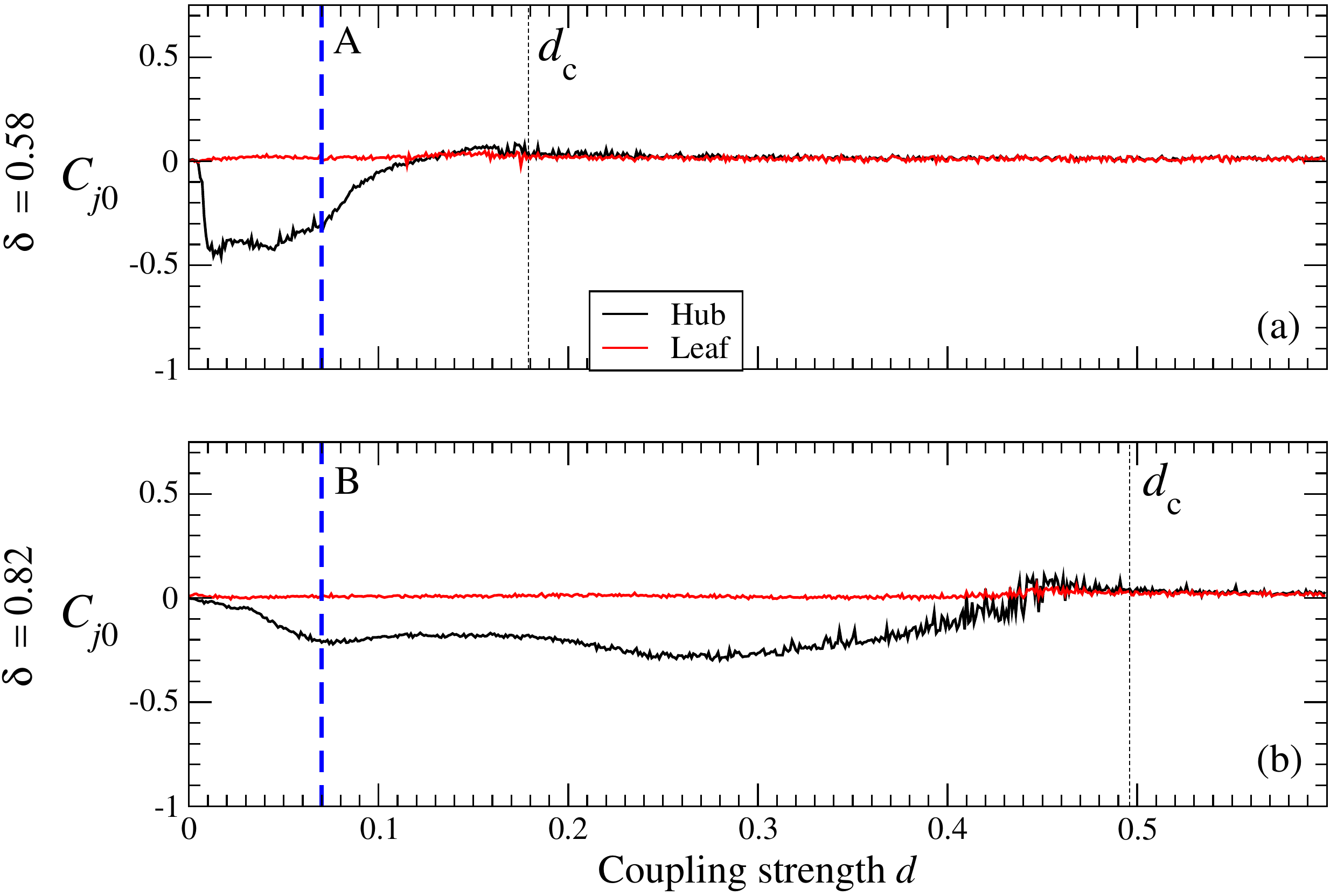} \\[0.2cm]
   \includegraphics[width=0.44\textwidth]{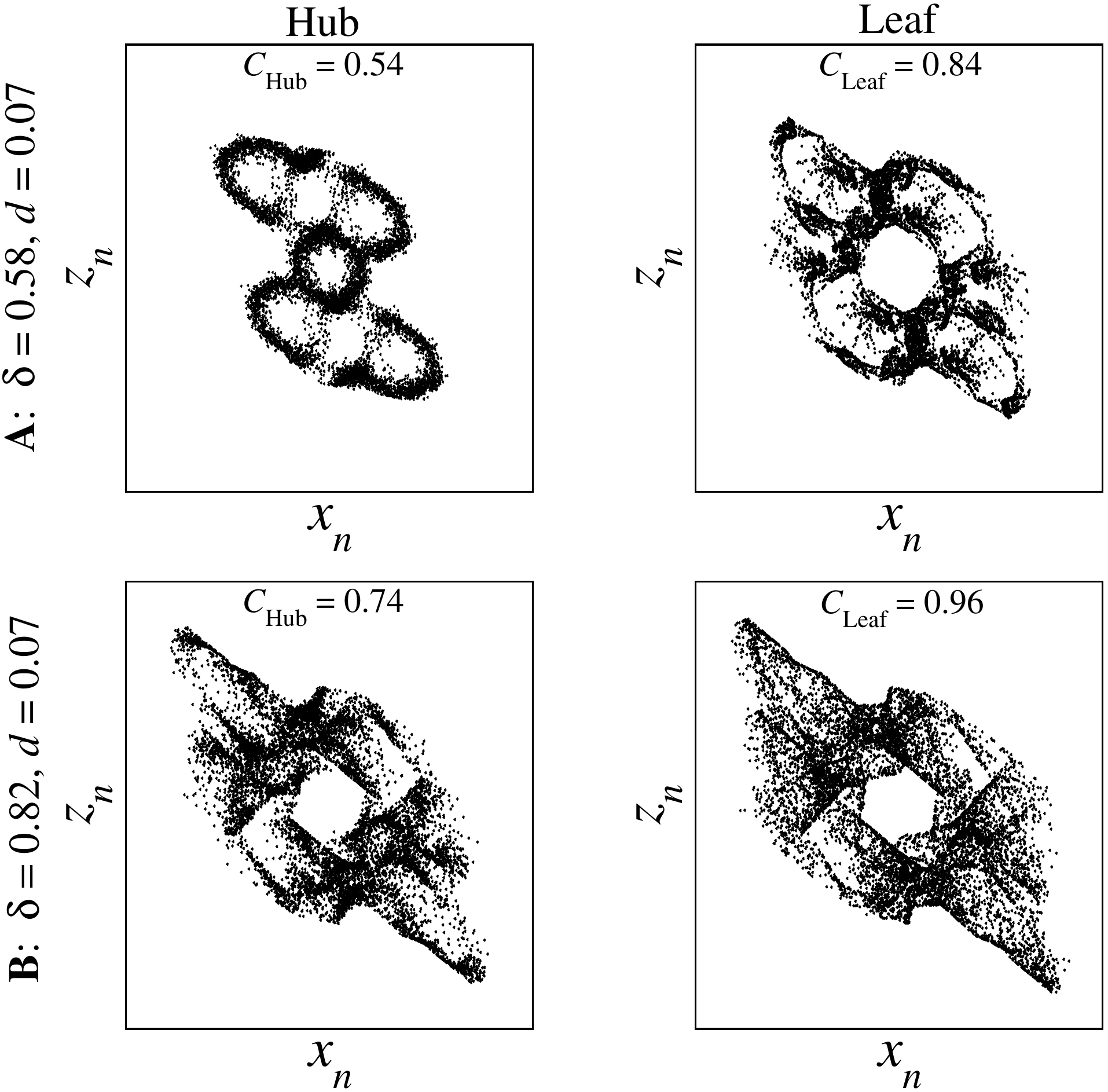} \\[-0.3cm]
   \caption{Relative complexity $C_{k0}$ as a function of the coupling strength 
$d$ of a star network with $N = 16$ $y$-coupled Saito models (\ref{saiteq}) for 
 (a) $\delta=0.58$ and (b) $\delta=0.82$. 
Poincar\'e sections for the hub and a leaf for (A) $\delta =0.58$ 
and for (B) $\delta = 0.82$, both with $d=0.07$.
Other parameter values: $\rho = 14$, $\eta = 1$, and $\epsilon = 0.01$. 
}
   \label{fig5Saito}
\end{figure}

In this section, to confirm the generality of the nodal dynamical differentiation route, we will investigate a completely different model of nonlinear dynamics, the Saito model,\cite{Sai90} which is able to generate a very 
rich dynamical behavior including toroidal  chaos. The vector flow in Eq.~(\ref{eq:dynet}) reads
\begin{equation}
  \label{saiteq}
 \mb{f}(\mb{x})=\left [\rho (-y +z), x + 2 \delta y, -x -w , (z - \Phi (w))/\epsilon\right]
\end{equation}
where $\mb{x}=(x,y,z,w)$ and 
\begin{equation}
    \Phi(w) = 
    \left|
      \begin{array}{ccl}
        w - (1 + \eta)  & & w \geq \eta \\[0.1cm]
        \displaystyle
         - \frac{w}{\eta}  & \mbox{ if } & |w| < \eta \\[0.3cm]
        w + (1 + \eta)  & & w \leq - \eta  \, . 
      \end{array}
    \right.
\end{equation}
This is a four-dimensional system involving a linear piecewise function as a switch mechanism. The dynamics 
produced by this model can be chaotic, quasi-periodic, toroidal chaotic, or even hyperchaotic, also structured 
around a torus. Parameters $\rho$, $\eta$, and $\epsilon$ are fixed and $\delta$ is used as a 
bifurcation parameter. Here we chose $\delta=0.58$ for toroidal chaos, and $\delta=82$ to 
produce hyperchaotic toroidal chaos. Close examples of the Poincar\'e sections of these attractors can be 
grasped, respectively, in Fig. \ref{fig5Saito}(A) and \ref{fig5Saito}(B) for a leaf of a star of $N=16$ 
$y$-coupled Saito models whose dynamics are almost identical to the uncoupled scenarios. The main difference 
between these two dynamics is the  ``thickness'' of the Poincar\'e section being much thiner for 
$\delta = 0.58$ than for $\delta = 0.82$. The hyperchaotic nature is revealed  by the overlapping structures of 
the Poincar\'e section as observed in the folded-towel map introduced by R\"ossler.\cite{Ros79d} This is 
partly confirmed with the Lyapunov exponents which for $\delta=0.58$ are
\[ \lambda_1 = 0.047 > \lambda_2 = 0.012 \approx \lambda_3 = -0.022 
   > \lambda_4 = - 94.79 \, , \]
with two null exponents as expected for toroidal chaos structured around a torus T$^2$,\cite{Let20b} and for 
$\delta=0.82$ are 
\[ \lambda_1 = 0.164 > \lambda_2 = 0.069 \approx \lambda_3 = -0.033 
   > \lambda_4 = - 94.70 \, , \]
with two positive and one null exponents as needed for toroidal hyperchaos. Note that in the latter case it is still unclear whether
the second positive exponent is merged or not with the third null exponent and this is an issue currently under study.
 Investigations, which are out of the scope of the present work, would allow determining whether the fourth dimension is required for embedding the toroidal chaos ($\delta = 0.58$).
Being hyperchaotic for $\delta = 0.82$, the dynamics are necessarily  four-dimensional.

When these models are coupled through the variable $y$, they synchronize as shown in 
Figs. \ref{fig5Saito}(a) and \ref{fig5Saito}(b) with the convergence of their relative complexity to zero, 
being  the critical coupling $d_{\rm c }$ larger for the hyperchaotic dynamics. Nevertheless, in both cases, what is preserved along the route to synchronization with these toroidal chaotic and hyperchaotic dynamics 
is the node differentiation mainly of the hub whose dynamics turns less developed  than the uncoupled one while the leaves sustain it over the whole range of coupling strengths [compare the first-return maps in Fig.~\ref{fig5Saito}(A), and in Fig.~\ref{fig5Saito}(B)]. The route to  synchronization appears very simple, 
most likely due to the constrained toroidal structure of the nominal dynamics.

\section{Networks of R\"ossler oscillators}
\label{sec:largernets}

Having characterized the relationship between the degree centrality and 
dynamical complexity in star networks, we move forward to generalize our 
results to networks with a broader degree distribution. This issue was 
partially tackled in Refs.\cite{Tla19,Min19}, where a strong correlation between 
degree $k$ and complexity $C_D$ allowed establishing a node 
hierarchy. However, given the present novel 
results revealing the important role played by the nominal nodal dynamics, we
extend our present study to larger networks.  

\begin{figure}[t]
  \centering
  \includegraphics[width=0.49\textwidth]{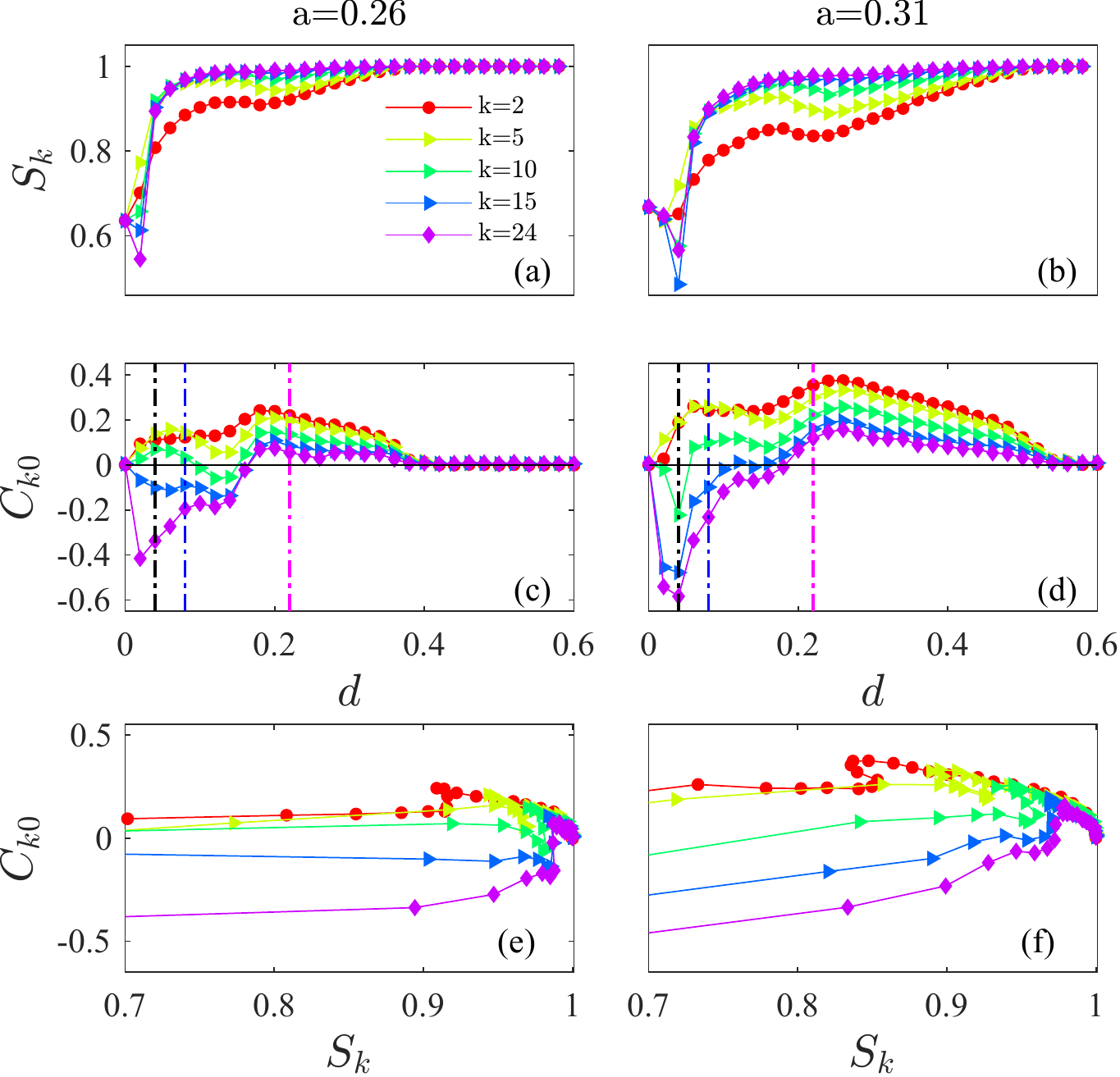} \\[-0.2cm]
  \caption{ (a,b) Average $k$-class $S_k$ phase 
synchronization parameter for several values of degree $k$ for $N=300$, 
$\langle k \rangle=4$ SF networks of $y$-coupled R\"ossler systems when 
$a=0.26$ (left panels) and  $a=0.31$ (right panels), (c,d) averaged relative 
complexity $C_{k0}$. The vertical dotted lines mark the couplings 
analyzed in Fig.\ \ref{fig:networks_ERSF} the lower panels:  $d=0.04$ (black), 
$d=0.08$ (blue) and $d = 0.22$ (magenta). (e,f) Relative complexity for 
different $k$-classes as a function of the phase synchronization $S_k$. \label{fig:networks-SF}}
\end{figure}

We maximize the degree heterogeneity by using Barabasi-Albert scale-free (SF)
networks of $N$ identical $y$-coupled R\"ossler oscillators retaining the parameter settings
provided in Section \ref{sec:StarRoss}. Since we expect that nodes having the 
same degree $k$ play equivalent roles in the network, we calculate the 
evolution of $C_k$ within a degree class $k$ by averaging over the $N_k$ nodes 
having degree $k$, that is, 
\begin{equation}
  C_k  = \frac{1}{N_k} \sum_{[j|k_j=k]} C_j \, ,  
\end{equation}
where $C_j$ is the dynamical complexity of the $j$th node. Here, we use the relative complexity 
$C_{k0} =  C_k - C_0$ which helps to better assess 
the effects of both the coupling and the topology in the complexity of the 
$k$-class nodes. In addition, to evaluate the impact of the nodal environment, 
we perform our calculations for networks made of $N= 300$ nodes, wherein
$\langle k\rangle=4$. All the results are averaged over 10 different networks 
realizations.  

First, in Fig. \ref{fig:networks-SF}(a-b) we plot 
the time averaged phase synchronization $S_k$ of $k$-class nodes with respect to the phase of the mean field for $N=300$ SF networks as a function of the 
coupling $d$. We define $S_k$  as 
\begin{equation}
S_k  = \frac{1}{N_k} 
   \displaystyle \sum_{{[j|k_j=k]}}^N S_j\, . 
\end{equation}
where $S_j$ is the time  averaged phase synchronization of node $j$ with the mean-field global phase, defined 
in Eq.~(\ref{S_j}). The result is then ensemble averaged  
over 10 network realizations. Along the route to synchronization, the $k$-classes synchronize hierarchically 
to the mean field before all of them lock at the critical coupling 
($d_{\rm c} = 0.37$ for $a = 0.26$ and $d_{\rm c} = 0.52$ for
$a = 0.31$),\cite{Zho06,Per10,Tla19} therefore existing node differentiation also in heterogeneous networks. While the phase synchronization $S_k$ is monotonously 
increasing for most of the coupling range and classes, the weakly coupled regime presents anomalous synchronization\cite{Bla03,Boa18} over which most of the $S_k$ are below the basal, uncoupled level. This anomalous range, more prominent for $a=0.31$, is associated with the maximal node differentiation [compare Fig.\ \ref{fig:networks-SF}(b) and \ref{fig:networks-SF}(d)]. 

As already observed in Section \ref{sec:StarRoss}, the less developed the
dynamics, the smaller the critical value $d_{\rm c}$ [Fig.\ 
\ref{fig:networks-SF}(c)-(d)]. In the weakly coupled regime, the relative complexity $C_{k0}$  shows a strong node differentiation: while the hubs present a markedly reduced complexity with respect to the nominal value (with a clear negative minimal), for the smaller degrees the complexity increases. This increment in the less connected nodes ($k = 2$ and $k = 5$ in the example) is non-monotonous with $d$ and $k$, as observed in the stars in Sections \ref{sec:StarRoss} and \ref{sec:StarOther}.

For stronger coupling all the nodes increase their complexity well above the nominal value, ordered following the reverse degree ranking [Fig. \ref{fig:networks-SF}(c)-(d)], recovering the scenario observed for the hub in Section \ref{sec:StarRoss}. All these features are qualitatively shared for the two different $a$-values, but the deviations from the uncoupled value are larger for the more developed chaos, $a=0.31$. 

Plotting the relative complexity $C_{k0}$ as a function of the phase synchronization 
$S_k$ reveals a dependency which is stronger for nodes with a larger
degree [Fig. \ref{fig:networks-SF}(e)-(f)]. Typically, low-degree nodes exhibit a dynamics which is nearly 
independent of the synchronization while it is the opposite for large-degree nodes. It also clearly shows that 
the relative complexity converges to zero for larger $d$-values when $k$ increases. This delineates another 
signature  of node differentiation. 

\begin{figure}[t]
  \centering
  \includegraphics[width=0.48\textwidth]{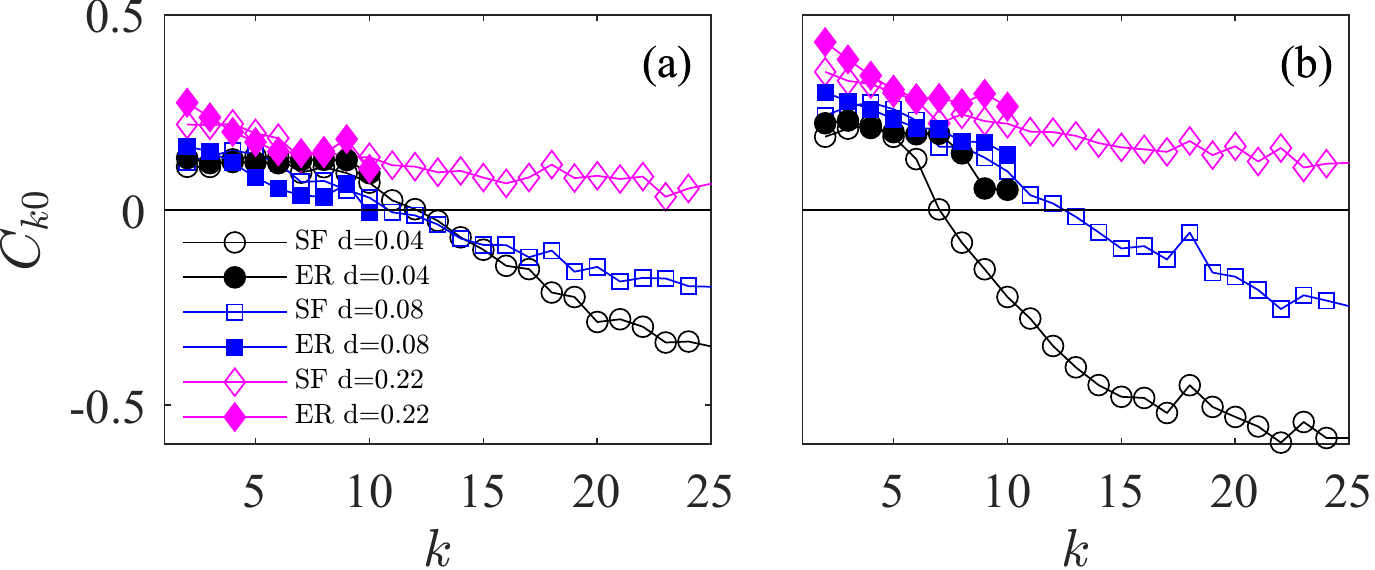} 
	\\[-0.3cm]
  \caption{Relative complexity $C_{k0}$ as a function of the degree $k$ for (a) $a=0.26$, and (b) $a=0.31$. In each panel, curves correspond to SF (void symbols) and ER (full symbols) with the coupling strengths marked in Fig.\ref{fig:networks_ERSF}(c,d) with vertical lines: $d = 0.04$ (circles), $d = 0.08$ (squares), and $d = 0.22$ (triangles). Results are averaged over 10 different 
 network realizations of $N=300$. Other parameter values as in Fig.\ \ref{fig1}.} 
  \label{fig:networks_ERSF}
\end{figure}

Therefore, we conclude that in most of the regimes it is possible to correlate the node degree with the relative complexity. Furthermore, the degree centrality is the single structural parameter that affects the node behaviour, while the rest of environmental topological features has no impact. This is shown in Fig. \ref{fig:networks_ERSF}, where we plot the value of the 
relative complexity $C_{k0}$ as a function of $k$ for three representative 
values of $d$, both for SF networks and ER networks with $\langle k \rangle =4$. The ER and SF curves overlap and, therefore, the dependence of $C_k$ on $a$ and $d$ are the same regardless of the topology. This is quite remarkable since the ER and SF networks have a 
different critical coupling $d_{\rm c}$: for a given $d$-value, their global 
dynamics are different, but the nodes of degree $k$ have equivalent dynamics 
independently of the environment. The same result is obtained for different 
sizes of both ER and SF networks (not shown). 

\begin{figure}[ht]
  \centering
  \includegraphics[width=0.48\textwidth]{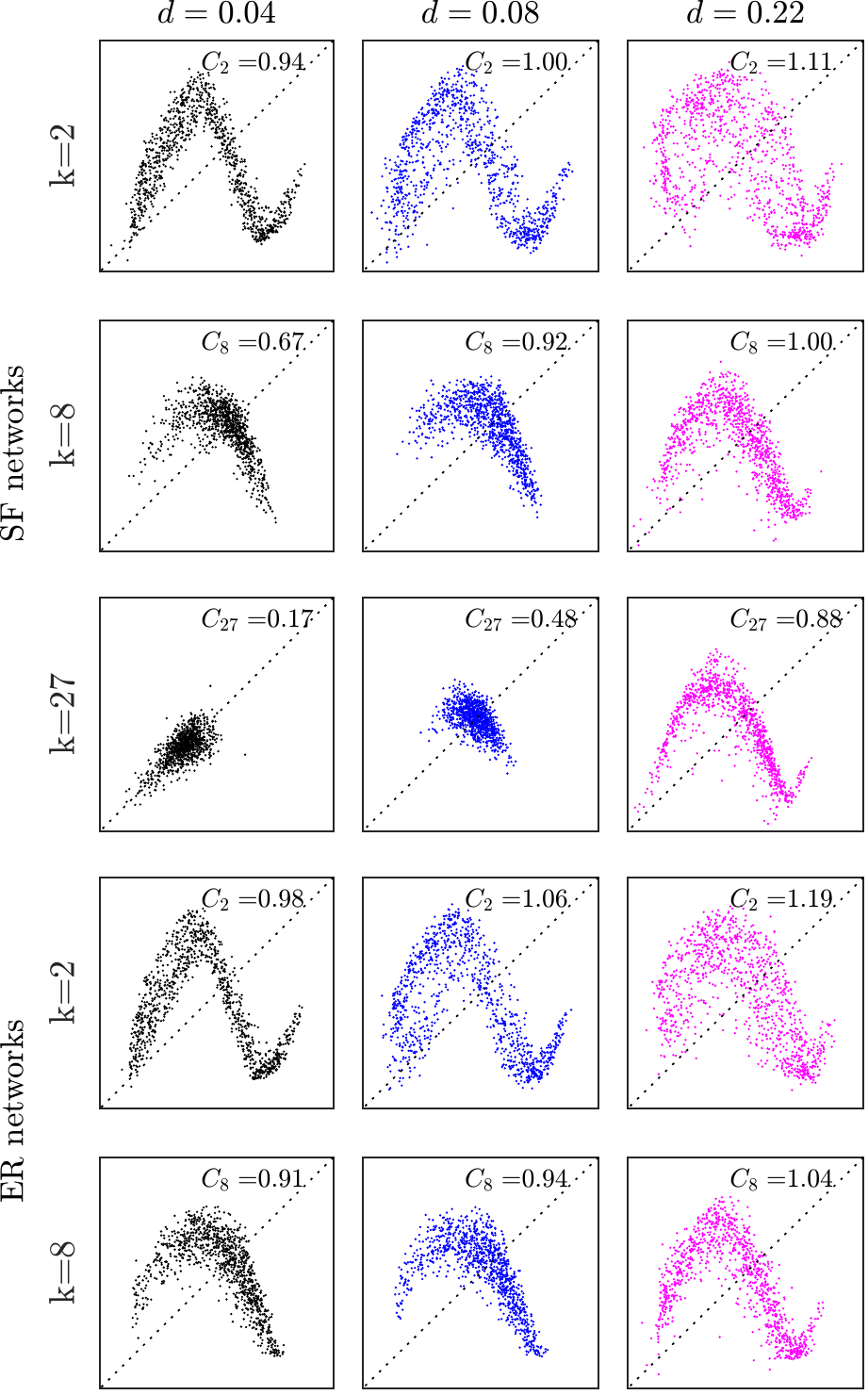} \\[-0.2cm]
  \caption{Different first-return maps to a Poincar\'e section for three 
$k$-classes of nodes along a route to synchronization in a $N=300$ 
($\langle k \rangle=4$) SF network of $y$-coupled R\"ossler nodes ($a= 0.31$ 
and other parameters as in Fig.\ \ref{fig1}). The complexity $C_{\rm D}$ is 
reported in each case.}
  \label{fig:SFreturnmaps}
\end{figure}

To better illustrate the node differentiation in larger networks, we plotted 
the first-return maps for three different $k$-classes of nodes along the route 
to synchronization (Fig. \ref{fig:SFreturnmaps}). Low-degree nodes ($k=2$)
in SF networks produce first-return maps whose thickness increases with the coupling strength,
up to $d < d_{\rm c}$ (top row in Fig.\ \ref{fig:SFreturnmaps}): for these 
nodes, the relative complexity is always positive, and the first maps resemble the map of the uncoupled dynamics but thicker (compare with the map for $a = 0.31$ in Fig.\ \ref{fig1}). 
For large degree nodes (third row in Fig.\ \ref{fig:SFreturnmaps}), once the minimum complexity is reached in the weakly coupled regime, the complexity increases with the coupling strength $d$. 
Around the minimum, the maps comprise a small cloud of points in the neighborhood of the inner singular point (bottom left
of the first-return map); this is progressively transformed into a ``noisy'' 
period-1 limit cycle, which is characterized by a cloud of points elongated 
perpendicularly to the bisecting line and located around the centre of the map
($d=0.08$ with $C_{27} = 0.48$). Before the onset of  
synchronization, large degree nodes produce a map which resembles the 
nominal one but is slightly thicker ($d=0.22$ with $C_{27} = 0.88$). The nodes with an intermediary degree ($k = 8$ in the example) produce a map which has features intermediate between the two extreme cases previously discussed: the node differentiation is, thus, evidently 
correlated with the node degree. As previously discussed, in large networks, 
node differentiation is mostly a monotonous function of the coupling 
strength and of the degree. When SF networks are replaced by ER ones, the degree range
is narrower, but the maps for $k=2$ and $k=8$ are very similar to their counterparts in SF networks.  We conclude that node degree is clearly the most important factor
for the node differentiation along the route to synchronization.

\section{Conclusion}
\label{conc}

Routes to synchronization need to be elucidated to attain a better understanding
and knowledge of the possible scenarios that may be encountered depending
on the coupling, nominal node dynamics, topology, and network size. Here, we confirmed and extended previous work depicting a non-trivial effect of connectedness on node dynamics, particularly the existence of a non-monotonic relationship between the complexity of node dynamics and coupling strength. There is, indeed, a node differentiation that evolves with
the coupling strength. Typically, when the coupling function provides 
type-{\sc i} synchronizability, low coupling strengths induce a significant
reduction in the complexity of the large-degree nodes, while those having a 
small degree are left nearly unaffected. Consequently, increasing the $d$-value, all the
nodes in small networks or those with a large degree in large networks present
a minimal dynamical complexity. In every network, all nodes have 
a complexity that increases, often reaching a greater level than the 
nominal one, before the onset of full synchronization. This sketch for the 
route to synchronization is clearly observed for the two three-dimensional
systems (R\"ossler and Lorenz). With more complex node dynamics (Mackey-Glass 
and Saito), the decrease towards a minimal complexity is only observed in the 
hub of star networks; further studies with other network types are still 
needed for attaining a more general view of the latter system. The node 
differentiation is not a monotonic function of the coupling nor of the 
synchrony. When the coupling provides a type-{\sc ii} synchronizability, the 
route to synchronization is an abridged version of the route observed with a
type-{\sc i} synchronizability.

\section*{Acknowledgments}
ISN and IL acknowledge financial support from the Ministerio de Econom\'ia, 
Industria y Competitividad of Spain under project FIS2017-84151-P.

\bibliography{SysDyn}
\end{document}